\PassOptionsToPackage{dvipsnames}{xcolor}
\PassOptionsToPackage{hyphens}{url}

\documentclass[conference]{IEEEtran}
\IEEEoverridecommandlockouts

\usepackage{_cfg-paper}

\newboolean{makeXTRA}
\setboolean{makeXTRA}{false} % default
\newcommand{\sysname}{\textsc{MemPower}}
\newcommand{\pstate}{CPU Power State}
\newcommand{\pstates}{CPU Power States}

\newcommand{\boldpara}[1]{\paragraph*{\textbf{#1}}}

\newcounter{goal}
\newcommand{\goal}[2]{%
  \refstepcounter{goal}%
  \noindent\textbf{Goal~G\thegoal: #1.} #2\par
}

\newcommand{\nrt}[1]{\begingroup\relax\ifmmode\boldmath\else\bfseries\sffamily\fi\color{Red}\ignorespaces(nrt)#1\ignorespaces\endgroup}

\ifthenelse{\boolean{makeXTRA}}
  {\newcommand{\XTRA}[1]{\phantom{}\begingroup\slshape\color{RoyalBlue}\ignorespaces#1\ignorespaces\endgroup}}
  {\newcommand{\XTRA}[1]{}}

\Crefname{section}{Section}{Sections}
\crefname{section}{\S}{\S}
\crefformat{section}{#2\S#1#3}

\Crefname{figure}{Figure}{Figures}% Crefname is default
\crefname{figure}{Fig.}{Figs.}

\Crefname{equation}{Equation}{Equations}
\crefname{equation}{Eq.}{Eqs.}

\Crefname{algocf}{Algorithm}{Algorithms}
\crefname{algocf}{alg.}{algs.}

\definecolor{shadecolor}{gray}{0.9} % framed

\usepackage[ruled,linesnumbered]{algorithm2e} % Required for typesetting algorithms
\usepackage{minted}

\definecolor{codegreen}{rgb}{0,0.6,0}
\definecolor{codegray}{rgb}{0.5,0.5,0.5}
\definecolor{codepurple}{rgb}{0.58,0,0.82}
\definecolor{hilightcolor}{rgb}{0.47,0.87,0.55}
\definecolor{backcolour}{rgb}{0.98,0.98,0.98}
\definecolor{keywordcolor}{rgb}{0.000000, 0.000000, 0.635294}

\lstdefinestyle{mystyle}{
    language=C++,
    backgroundcolor=\color{backcolour},
    commentstyle=\color{codegreen}\ttfamily,
    keywordstyle=\color{keywordcolor}\ttfamily,
    numbers=left,
    numberstyle=\tiny\color{codegray},
    stringstyle=\color{codepurple},
    basicstyle=\footnotesize\ttfamily,
    breakatwhitespace=false,
    breaklines=true,
    captionpos=b,
    keepspaces=true,
    numbers=left,
    numbersep=5pt,
    rulecolor=\color{black},
    frame=tb,
    showspaces=false,
    showstringspaces=false,
    showtabs=false,
    tabsize=2
}
\newcommand{\mylstset}{%
  \mylstsetCommon{}%
  \lstset{%
  }
}

\mylstset{}

\lstdefinestyle{ics-mono}{
  language=C,
  basicstyle=\ttfamily\normalsize,
  numbers=left,
  numberstyle=\ttfamily\scriptsize\color{gray},
  stepnumber=1,
  numbersep=8pt,
  showstringspaces=false,
  breaklines=true,
  columns=fullflexible,
  keepspaces=true,
  commentstyle=\color{gray!70},         % monochrome comments
  keywordstyle=\bfseries,            % bold “keywords”
  keywordstyle=[2]\bfseries\underline,
  morekeywords={int,struct,return},
  morekeywords=[2]{mp_set_pstate_min, mp_set_pstate_max, mp_set_pstate_eff, mempow_init, mempow_deinit},
  frame=none,
  xleftmargin=2em,
  xrightmargin=2em,
}

\makeatletter
\newcommand{\linebreakand}{%
  \end{@IEEEauthorhalign}
  \hfill\mbox{}\par
  \mbox{}\hfill\begin{@IEEEauthorhalign}
}
\makeatother

\begin{document}

%============================================================================
% title
%============================================================================

\title{
\sysname{}: Efficient Power Management with Fine-grained Memory Analysis and Modeling for HPC Workloads
\thanks{This research is supported by the U.S.\@ Department of Energy (DOE) through
  the Office of Advanced Scientific Computing Research's
``Advanced Memory to Support Artificial Intelligence for Science'' (76125)
and 
``Orchestration for Distributed \& Data-Intensive Scientific Exploration'' (77765).
Pacific Northwest National Laboratory
is operated by Battelle for the DOE under Contract DE-AC05-76RL01830. Additional support is provided by the U.S. National Science Foundation (NSF) via awards CNS-2533773 and CNS-2310422.}
}

\author{%
  \IEEEauthorblockN{Nanda Velugoti}
  \IEEEauthorblockA{Oregon State University\\
    velugotn@oregonstate.edu%
  }
  \and
  \IEEEauthorblockN{Joseph Manzano}
  \IEEEauthorblockA{Pacific Northwest National Laboratory\\
    joseph.manzano@pnnl.gov
  }
  \and
  \IEEEauthorblockN{Andres Marquez}
  \IEEEauthorblockA{Pacific Northwest National Laboratory\\
    andres.marquez@pnnl.gov
  }
  \linebreakand
  \IEEEauthorblockN{Nathan Tallent}
  \IEEEauthorblockA{Pacific Northwest National Laboratory\\
    nathan.tallent@pnnl.gov
  }
  \and
  \IEEEauthorblockN{Kyle C. Hale}
  \IEEEauthorblockA{Oregon State University\\
    kyle.hale@oregonstate.edu
  }
}

\ifthenelse{\boolean{maketitleAfterAllMeta}}{}{\maketitle}

%============================================================================
% abstract
%============================================================================

\begin{abstract}
Managing the energy consumption and power efficiency of parallel applications
is a significant issue in both HPC environments and in the cloud. As
emerging applications continue to push against the memory wall of modern
machines, the growing imbalance between compute and data movement creates
new opportunities to intelligently tune CPU power consumption.
Unfortunately, existing frequency and voltage scaling techniques do not
adequately capture fine-grained changes in memory access behavior,
rendering the compute/data access imbalance invisible to the components of
the system that could capitalize on it, thus leaving potential power
savings on the table.   

In this paper, we propose \sysname{}, a flexible, model-based approach to
exposing compute/data movement imbalance that characterizes the
fine-grained memory behavior of parallel workloads.
This characterization then informs our automated software framework which
can statically instrument the application binary with model-determined
voltage/frequency transitions that balance fine-grained changes in memory
access behavior with the costs of hardware transitions. 
Using \sysname{}, we demonstrate a reduction in EDP of 6\% to 42\% on a range
of HPC benchmarks with minimal impact on execution time when compared to
the standard OS/hardware-managed power control mechanism. 

\begin{IEEEkeywords}
memory trace analysis, energy savings, power management, cost model, binary instrumentation, Linux kernel
\end{IEEEkeywords}

\end{abstract}

%===========================================================

\ifthenelse{\boolean{maketitleAfterAllMeta}}{\maketitle}{}

% Insert IEEE copyright notice as a floating bottom banner
\begin{tikzpicture}[remember picture, overlay]
  \node[anchor=south, yshift=10pt] at (current page.south) {%
    \parbox{\textwidth}{%
      \footnotesize \centering
      \textcopyright~2026 IEEE. Personal use of this material is permitted. Permission from IEEE must be obtained for all other uses, in any current or future media, including reprinting/republishing this material for advertising or promotional purposes, creating new collective works, for resale or redistribution to servers or lists, or reuse of any copyrighted component of this work in other works.
    }%
  };
\end{tikzpicture}

%============================================================================
%============================================================================

\section{Introduction}
\label{sec:introduction}

The impact of CPU power management on overall energy efficiency in the context
of HPC applications is well
known~\cite{Ilsche:2024:CLUSTER:energy-insights-HPC, Zhang:2016:ASPLOS:PUPIL},
and modern systems can effectively reduce power using hardware- and OS-managed
schemes. These systems all rely on some flavor of dynamic voltage and frequency
scaling (DVFS), where the CPU clock rate and voltage are adjusted to modulate
performance per Watt~\cite{WEISER:1994:DVFS}. Early DVFS schemes such as  Intel
SpeedStep~\cite{SPEEDSTEP} allowed the OS to manage voltage and frequency
using a hardware-exposed interface, for example with Linux
CPUFreq~\cite{LINUX-CPUFREQ}. However, more recent Intel chips (since Skylake)
have shifted DVFS control back to the hardware via SpeedShift, with the stated
goal of mitigating software latency and allowing more responsiveness to
instantaneous microarchitectural events~\cite{DOWECK:2017:SKYLAKE}. In this regime,
software can provide hints, and the hardware can respond to them appropriately
in conjunction with its own collected metrics. With increasing hardware
control, and hints that machine learning models will be tightly integrated in
the hardware power control loop for coming
chips~\cite{PIERSMA:2025:DVFS-REVERSE}, it is paramount to ensure that the
hardware has sufficient information to effectively modulate energy efficiency.

\begin{figure}
\centering
\includegraphics[width=0.8\columnwidth]{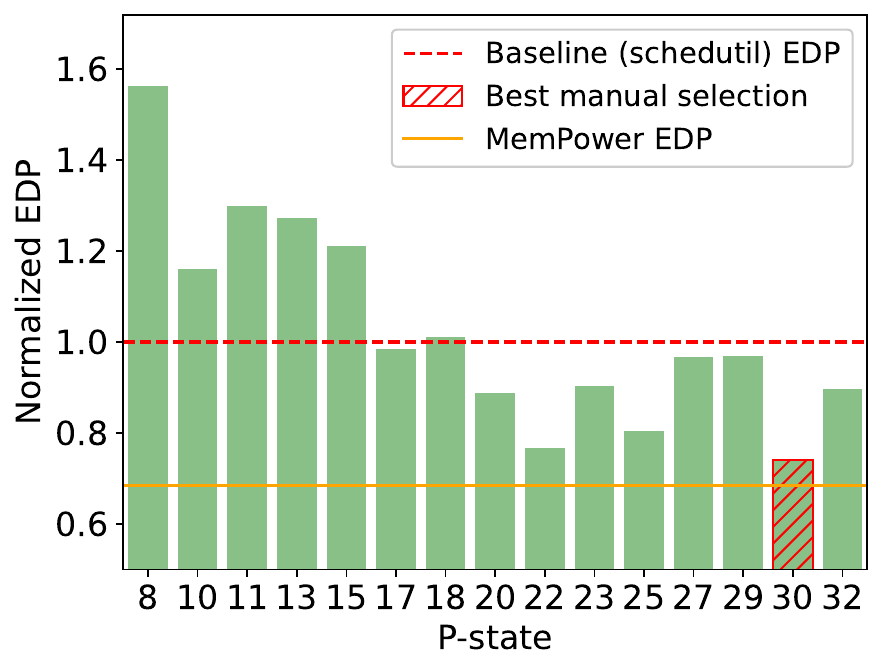}
\caption{Relative energy-delay product (EDP) of \textit{miniVite} when executed
at all \pstate{} configurations. \sysname{} outperforms both the manual
power selection scheme and the OS baseline (\texttt{schedutil}). Lower is
better.}
\label{fig:motivation}
\end{figure}

The state-of-the-art automatic power management scheme,
schedutil~\cite{LINUX-SCHEDUTIL}, performs well on memory intensive
applications that are bandwidth bound, as it analyzes the workload's
instruction mix on demand to select an effective \pstate{}. However, it does
not adequately exploit the opportunities in memory intensive applications that
are latency-bound~\cite{GROPP:2000:CFD, KASHI:2025:HPGMXP}, as it lacks
information about the dynamic access trace. When a core is already stalled
waiting on memory, reducing its frequency costs little performance.  This means
that the more latency-bound an application is, the greater the opportunity
for transitioning into a lower \pstate{} to save energy with minimal
performance cost.

For example, consider the parallel graph workload, \textit{miniVite}~\cite{GHOSH:2018:PMBS:miniVite}, which is
known to have memory intensive code regions~\cite{Kilic:2022:CLUSTER:MemGaze}.
We modified this application so that memory intensive code regions can be
executed at manually controlled frequencies/voltages using software-directed DVFS (ACPI
P-states). Figure~\ref{fig:motivation} shows the energy delay product (EDP) for
this workload, capturing both energy consumption and execution time on
a modern Alder Lake machine. We compare the modern Linux approach for power
control (\texttt{schedutil}, discussed further in \S\ref{sec:background}) as baseline with
a simple approach that sets a static P-state for memory intensive functions for the entire run of the workload and \sysname{}
(described in \S\ref{sec:design-and-desc}).
Surprisingly, there are several \textit{fixed} DVFS states that do better than
the default dynamic OS/hardware coordination mechanism as seen in Figure
\ref{fig:motivation}, with P-state 30 (second highest \pstate{}) being the best setting (red hatched bar). 

The reason for this effect is that the baseline scheme treats all memory
accesses as equivalent and fails to distinguish between those with short and
long latencies. However, identifying code regions with latency-bound accesses for a given workload would either require access latencies to be exposed by hardware performance counters or would require detailed microarchitectural and cache simulation. Building on either approach runs the risk of sacrificing generality. 

Motivated by this observation, we propose a new model-based, memory-centric
approach to coordinating \pstates{} that focuses on capturing fine-grained memory-bound behavior in a way that generalizes across
workloads. Our approach does not require complex microarchitectural modeling or simulation to effectively characterize the workload
in terms of memory latencies. 
Software-directed DVFS is certainly not new~\cite{MAGLIKIS:2003:PROFILE-DVFS, WU:2005:JIT-DVFS,
CICOTTI:2013:ES, HOFFMANN:2011:DYN-KNOBS}, even at the compiler
level~\cite{XIE:2003:COMPILER-DVFS}. Others have attempted to bridge the
semantic gap between hardware, system software, and the
application~\cite{Wamhoff:2014:ATC:TURBO}, even with learned
models~\cite{MUKHERJEE:2025:CRAVE}. This includes leveraging phase behavior,
e.g., of distributed message passing~\cite{Cesarini:2018:ANDRE:COUNTDOWN,
Lim:2006:SC:DVFS-MPI} and the decoupling of data access and
execution~\cite{Jimborean:2014:CGO:Compiler-VFS, KOUKOS:2013:DAE-DVFS,
KOUKOS:2016:MV-DAE} for energy efficiency.  However, prior work is limited in
terms of capturing dynamic memory access patterns to make efficient and
granular power management decisions.

We propose augmenting the hardware/OS with \sysname{}, an analytical model paired with
a framework that transforms the target application binary into one that is more
energy-efficient. The goal of the model is to analyze the memory access
patterns of a given application at fine granularity (i.e., individual and architecturally visible loads
made by the application) and modulate the CPU clock frequency based on
memory behavior at runtime. Our key insight is that a model-based framework is uniquely
suited to extracting high-level information that can better capitalize on
memory-bound code regions compared to the hardware's available heuristics. 

\sysname{} builds on MemGaze~\cite{Kilic:2022:CLUSTER:MemGaze}, a low-level
memory trace analysis tool, to characterize the memory access patterns of code
regions of a target binary. To perform frequency management, we use Intel's
DVFS mechanism~\cite{INTEL_SYS_MAN} by implementing a userspace library and
a custom OS kernel driver (see \S\ref{sec:implementation}) for Linux.

We evaluate \sysname{} on a range of micro- and application benchmarks 
and demonstrate a reduction in overall EDP of 6\% to 42\%.  Our contributions
are as follows:

\begin{itemize}
    \item We investigate the idea of selecting a lower \pstate{} for memory intensive portions of an HPC application and empirically show that executing these portions at a single lower \pstate{} (i.e., manual selection) can result in better overall EDP compared to the baseline \verb|schedutil|.
    \item We propose \sysname{}, a model-based power management framework that profiles the dynamic memory behavior of a target binary and uses high-level memory metrics to extract memory intensive code regions. An analytical model then chooses efficient \pstate{} transitions for those regions, which \sysname{} injects into the binary to produce a more energy-efficient one.
    \item We evaluate \sysname{} on memory intensive HPC benchmarks to show that our model-based solution reduces the overall EDP by up to 42\% when compared to the baseline approach.
\end{itemize}

%============================================================================
%============================================================================

\section{Background}
\label{sec:background}
Here we provide background on hardware-managed dynamic voltage and frequency scaling, 
OS control of DVFS states, and the tools we build on. 

\boldpara{DVFS}
DVFS~\cite{WEISER:1994:DVFS} aims to manage a CPU's power consumption by
taking advantage of the superlinear savings one can achieve by changing
\textit{both} voltage and frequency\footnote{This is due to the switching
equation determining the dynamic power of a CPU, $P=\alpha C V^{2} f$.} at the
same time. This is achieved using circuits that modulate the CPU's supply
voltage and core clock frequencies. In the ACPI standard, each $(V, f)$ voltage–frequency pair for a processor defines an ``operational state,'' with a fixed number of such states determined by the vendor. Modern ACPI exposes these states (and more) via Collaborative Processor Performance Control (CPPC)~\cite{ACPI-SPEC-V6}, which higher software layers use to coordinate with hardware when selecting processor power states~\cite{LINUX-CPCC}. Policies for choosing operational states may be implemented in hardware, firmware (BIOS), or the OS, with recent microarchitectures favoring hardware control (e.g., Intel's Hardware-Controlled Performance States, HWP\footnote{See Section 16.4 of the Intel system programming guide~\cite{INTEL_SYS_MAN}.}). In all cases, vendors expose DVFS control via an extended control register set (MSRs on Intel and AMD). Under the older SpeedStep interface, software sets P-states directly~\cite{SPEEDSTEP}, while with HWP (introduced in Skylake) the OS provides ``hints” and the hardware ultimately selects the operational state. Throughout this paper we identify a \pstate{} by the frequency ratio (aka. P-state ranging from 8 till 32) written
to the \texttt{IA32\_PERF\_CTL} MSR rather than by its ACPI index, so larger
numbers denote higher frequency; note that this is the reverse of the ACPI
convention, in which $P_0$ is the highest-performance state.

\boldpara{Linux software control}
In the Linux kernel, \verb|intel-pstate| (for Intel CPUs) and \verb|acpi-cpufreq| 
(generic) are two kernel drivers that abstract away the platform-level hardware 
details associated with DVFS control. Each driver comes with a set of \textit{frequency governors} which integrate deeply with the
kernel's scheduler and provide finer-granularity information at lower latency
to direct the hardware~\cite{LINUX-SCHEDUTIL}. The \verb|intel-pstate| active mode has \verb|powersave| and \verb|performance| governors, whereas both passive mode and the
\verb|acpi-cpufreq| driver have access to extra governors: \verb|conservative|, \verb|ondemand|, \verb|userspace|, and \verb|schedutil|. We use the \verb|acpi-cpufreq| driver with the \verb|userspace| governor for manual instrumentation.

\boldpara{Software-driven DVFS}
We aim to use application information derived from our modeling framework to
drive \textit{direct} DVFS transitions from software. For
example, to switch from \pstate{} $P_i$ to $P_j$ manually, we can write a value
$P_j$ to a \textit{model-specific register} (MSR) of the target core (namely,
the \verb|IA32_PERF_CTL| MSR). After some delay (at least 6046
ns~\cite{Wamhoff:2014:ATC:TURBO}), the hardware will complete the transition to
a new frequency and/or voltage. Because this MSR is only accessible by
privileged software, a user program can only initiate these transitions by
delegating via the kernel interface. On Linux, this can be achieved with
mechanisms like system calls, \verb|ioctl|, \verb|sysfs| interface, and netlink 
sockets. In our case, the final MSR manipulation is performed in a custom kernel 
driver that exposes an interface to userspace. \S\ref{sec:pstate-impl} 
details both our userspace and kernel-resident components. 

\boldpara{MemGaze}
MemGaze~\cite{Kilic:2022:CLUSTER:MemGaze} is a low-level memory
analysis tool that dynamically captures memory access traces of an application
binary. Instead of capturing these at the software level,
MemGaze uses Intel's modern hardware tracing capability (enabled with the
\verb|PT_WRITE| instruction)~\cite{INTEL_SYS_MAN} to capture memory access
information during a program run. MemGaze instruments the target binary using
\verb|DynInst|~\cite{Bernat:2011:PASTE:DynInst} to inject \verb|PT_WRITE|
instruction before every memory access. It then runs the instrumented binary
to collect detailed memory traces, which inform various sophisticated analyses
that measure \textit{footprint growth rate} (i.e., average footprint per
memory access; see Table~\ref{tab:terminology}), \textit{reuse distance}, \textit{spatio-temporal locality}, and more.

\subsection{Identifying Profitable Power State Changes} % \pstate{}
\label{sec:pstate-challenges}

\begin{figure}
    \includegraphics[width=\columnwidth]{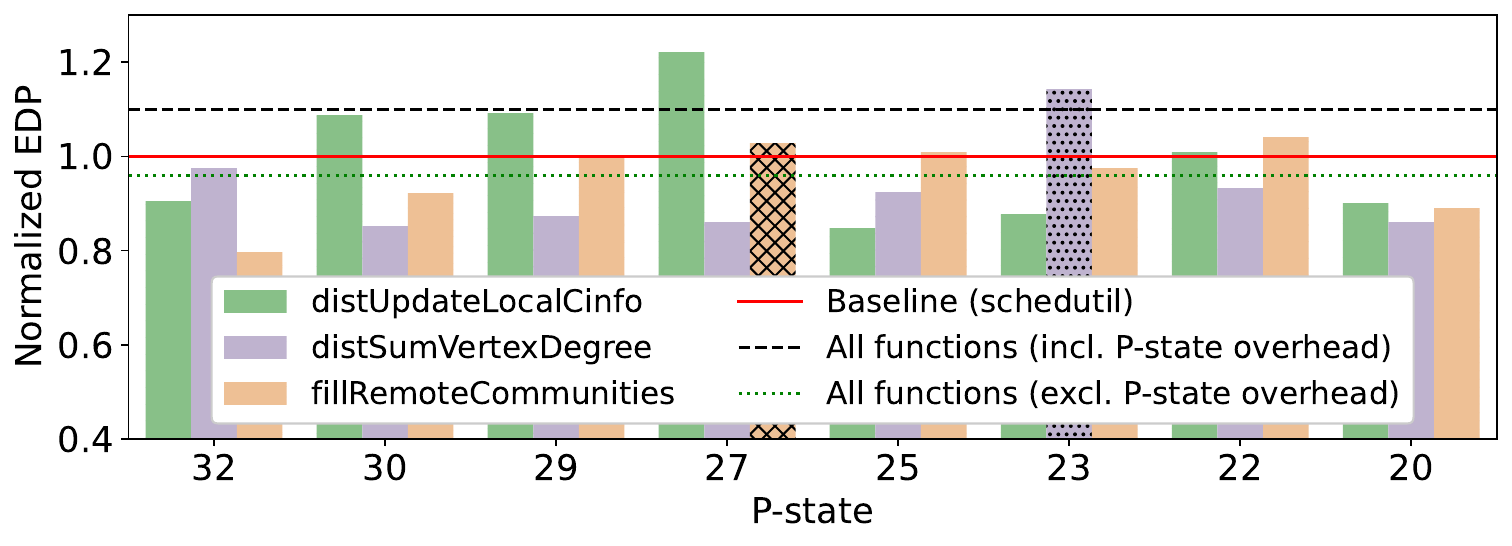}
            \caption{Illustrating the complexity of capturing the \pstate{} behavior and its effects on the EDP of an application (\textit{miniVite}). This shows that even when ideal P-states are selected for code regions, the overall EDP can be worse due to P-state transition overhead. Lower is better.}
    \label{fig:complexity}
\end{figure}

Designing a model that chooses an optimal \pstate{} setting for a given code
region using an aggregate performance metric is challenging as the relationship
between the \pstate{} setting and any type of memory behavior is complex. 
Furthermore, any model must account for the fact that state changes incur a switching cost that counteracts any benefit; these can be on the order of microseconds.
To illustrate this effect, we take the \textit{miniVite} workload, which clearly has
opportunity for better \pstate{} selection as shown in
Figure~\ref{fig:motivation}, and instrument three different code regions
(functions) individually: \verb|distUpdateLocalCinfo|, \verb|distSumVertexDegree|, and
\verb|fillRemoteCommunities|. We run the entire workload with one function instrumented at a time, at \pstates{} ranging from 20 to 32. We also run
the same application once more, only this time we select the ideal \pstate{}
for each function that resulted in the best EDP from previous runs, i.e., \pstates{} 25, 30, and 32 are chosen for \verb|distUpdateLocalCinfo|, \verb|distSumVertexDegree|, and \verb|fillRemoteCommunities| functions respectively.
Figure \ref{fig:complexity} shows the normalized EDP (Energy Delay Product) of
\textit{miniVite} with all of these configurations, and we observe
that, even though we empirically choose the ideal \pstates{} for each
individual function, the overall EDP (dashed line) is worse than the baseline
(\verb|schedutil|). 

The reason for this behavior is that the simple approach of picking the best (empirical) \pstate{} manually based on traces does not consider the overhead of placing a \pstate{} transition call within the code, especially when the transition overhead can quickly dominate the workload's runtime. For example, placing a transition inside a loop or inside a recursive function will incur the transition cost more than necessary. We also counted the transitions that occurred in order to estimate their performance overhead; Figure~\ref{fig:complexity} shows that if this overhead were avoided, the overall EDP (dotted line) would be lower than the baseline. Moreover, we consider this to be a conservative estimate of the EDP because it does not incorporate the energy savings obtained due to the reduced execution time of the program. 

\begin{figure}
    \centering
    \includegraphics[width=.8\columnwidth]{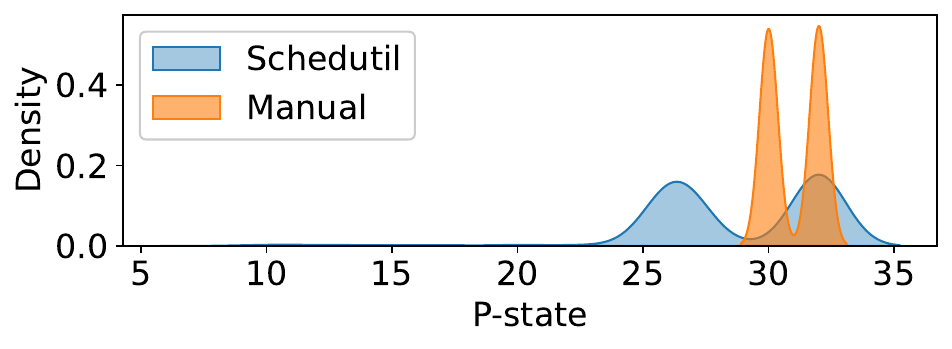}
    \caption{Manual vs. \textit{schedutil} \pstate{} distribution for \textit{miniVite}.} 
    \label{fig:pstate-dist}
    %\vspace{-0.2in} %% JBMF: Delete if more space can be found
\end{figure}

The effect motivates a more principled model that can statically predict advantageous (not necessarily optimal) power transitions and injection locations for a given workload based on its memory behavior such that the overall \pstate{} transition overhead is amortized over time. Moreover, Figure~\ref{fig:pstate-dist} compares the \pstates{} that are chosen by \verb|schedutil| (baseline) and manual (i.e., static/fixed) \pstate{} selection approaches (interpolated using KDE) for \textit{miniVite} workload and shows that \verb|schedutil| chooses suboptimal \pstates{} compared to manual selections which are the highest (32) and the ideal (30) \pstates{}. This explains why static \pstate{} setting outperforms \verb|schedutil| as described in Figure~\ref{fig:motivation}.

\section{\sysname{} Design}
\label{sec:model}

\sysname{} is a power management framework that finds efficient \pstate{} transitions fine-tuned for memory intensive code regions by profiling the target application's memory behavior and automatically injecting these transitions at the binary level to achieve good performance per Watt.

\subsection{Design Goals}
\label{sec:design-goals} 

The primary objective of \sysname{} is to capitalize
on memory-bound code regions to uncover power control
opportunities not available to the hardware/OS-coordinated default mechanism. We lay out the following goals in designing our framework: \\

\goal{Improved EDP}{
\sysname{} should improve energy-delay product compared to the baseline.
}
\label{goal:edp}
\

\goal{Identify latency-bound code regions}{
\sysname{} should capture irregular memory intensive access behavior.
}
\label{goal:irreg}
\

\goal{Balance}{
\sysname{} should balance energy utility and fidelity with cost amortization.
}
\label{goal:balance}
\

\goal{Minimal burden}{
To reduce the user's programming burden, \sysname{} must automatically instrument and manage the target application's power settings.
}
\label{goal:low-burden}
\

\boldpara{Assumptions} We assume that applications execute on Intel x86 CPUs (with \verb|PT_WRITE| support), running Linux, as \sysname{}'s \pstate{} management mechanism is implemented as a Linux kernel driver along with \verb|acpi-cpufreq| driver. Even though \sysname{} is application agnostic, we assume HPC workloads, focusing on applications at the node-level that leverage OpenMP for intra-node parallelism. 

We first discuss how to achieve \textbf{Goal G\ref{goal:irreg}}.

\subsection{Characterizing Memory and Power Behavior}
\label{sec:work-char}

Critical to \sysname{} is the ability to identify opportune regions of code to modulate \pstate{} based on memory-bound behavior. The underlying memory trace is provided by MemGaze. In order to choose an effective \pstate{} based on memory behavior, we first need  metrics that capture both the temporal reuse and operational intensity of a given code region within the target application. We achieve this using \textit{footprint growth rate} (i.e., average footprint per access) and \textit{access class} (i.e., memory accesses that are of type \textit{irregular} or \textit{constant} or \textit{strided}) where the former metric captures memory intensity and the latter captures locality.

The intuition is that if a particular code region displays irregular memory accesses that exhibit high footprint growth, then that code region is likely to be latency-bound due to stalls waiting on the memory system, and therefore a prospect for executing at a lower \pstate{}. In other words, if the core pipeline cannot extract sufficient instruction-level parallelism to hide DRAM accesses caused by cache pressure, the workload will be in a memory latency-bound rather than bandwidth-bound regime. Choosing a lower \pstate{} for code regions with poor locality is worthwhile because the existing power management scheme (\verb|schedutil|), which we primarily compare against, does not have such information about fine-grained memory accesses made by an application.

\begin{figure}
    \centering
    \includegraphics[width=\columnwidth]{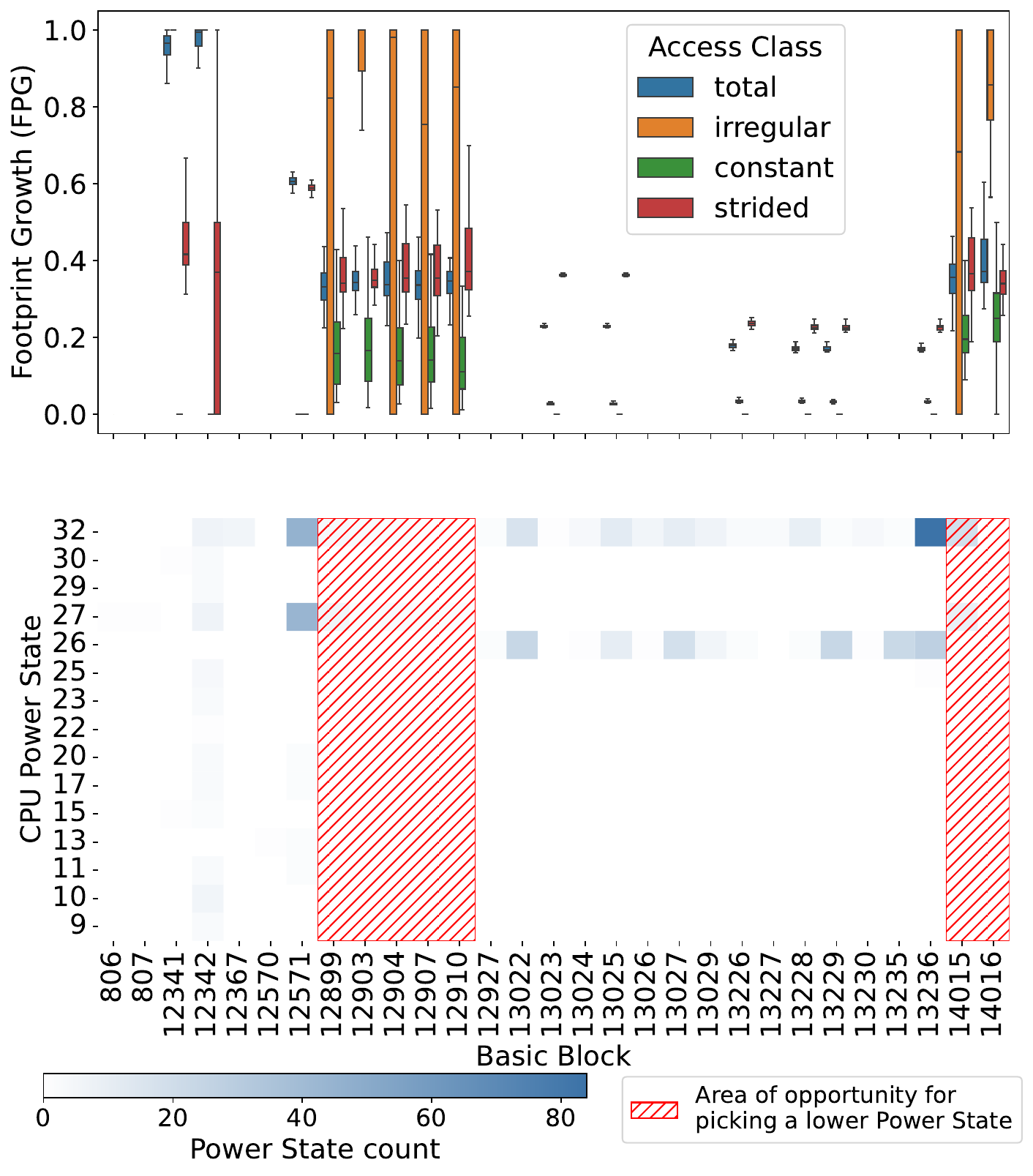}
    \caption{Areas of opportunity (red hatched rectangles) to pick an efficient (lower) \pstate{} for memory intensive code regions of \textit{miniVite} workload.} 
    \label{fig:fpg-vs-pstate}
\end{figure}

To identify areas of opportunity to improve energy consumption when compared to the baseline approach (\verb|schedutil|), we analyze the memory trace of \textit{miniVite} collected by running it through MemGaze and extract \textit{footprint growth rate}  categorized based on the \textit{access class}. We also use Linux's \verb|perf| tool to collect the application power trace to see the frequency distribution of \pstate{} transitions that occurred over time, directed by \verb|schedutil|. 
For both traces, we attribute the data to basic blocks in the application binary, which lets us correlate memory metrics with power transitions.

In Figure~\ref{fig:fpg-vs-pstate}, the top half shows the \textit{footprint growth rate} distribution per basic block based on the \textit{access class} for \textit{miniVite}. This is a distribution because we combine \textit{footprint growth rates} of all code regions that contain the corresponding basic block. The bottom half of Figure~\ref{fig:fpg-vs-pstate} shows the \pstate{} transition distribution per basic block of \textit{miniVite} represented as a heatmap, where the hue represents the number of transitions that have occurred for each available \pstate{}. Furthermore, we can also observe that there are clear areas of opportunity (highlighted with hatched rectangles on the heatmap) for picking a lower \pstate{} to save energy, as the code regions in those areas have high \textit{footprint growth rate} distribution of type \textit{irregular access} compared to \verb|schedutil|. In those regions, \verb|schedutil| used only two \pstates{}, allowing us to choose a lower \pstate{} to reduce energy.

\subsection{Model Design and Description}
\label{sec:design-and-desc}

\begin{figure}
\centering
\includegraphics[width=\columnwidth]{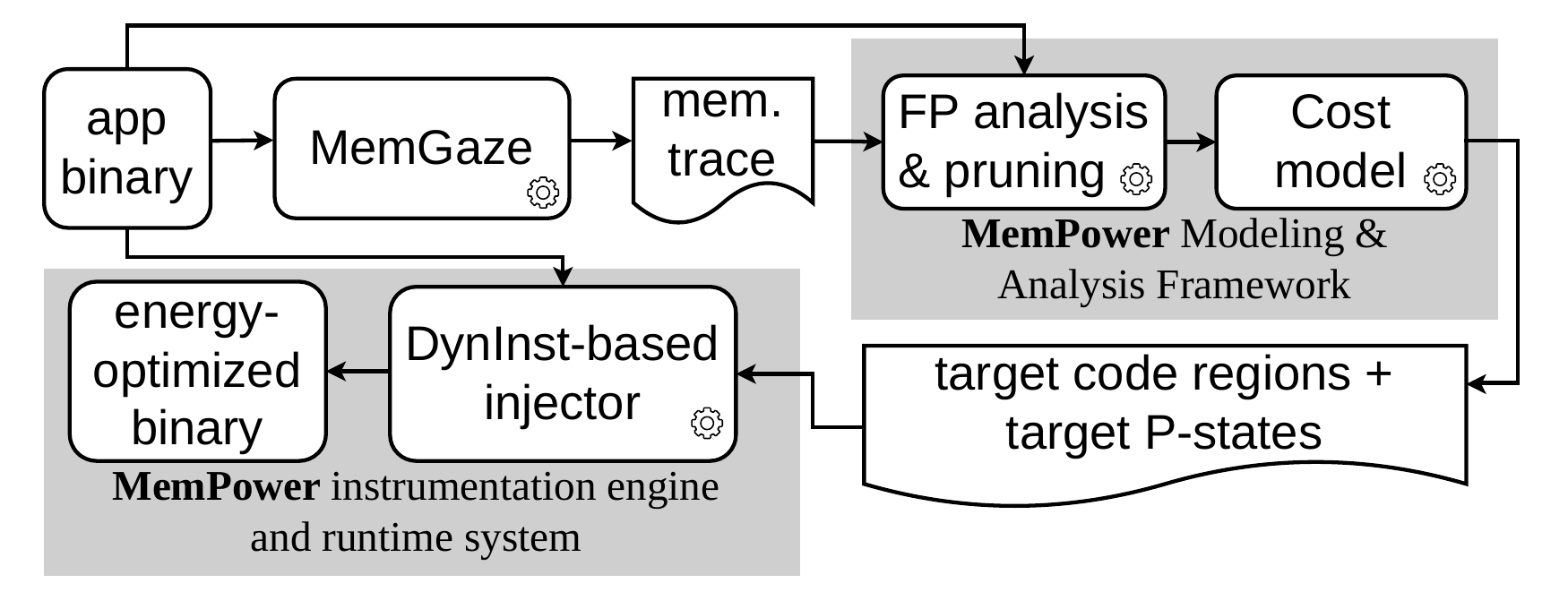}
\caption{High-level design of \sysname{}.}
\label{fig:model-high-level}
\end{figure}

At a high level, to choose effective \pstate{} transitions, we propose a model where we analyze the target program's memory access pattern to identify code regions with both irregular memory accesses and high footprint growth rate. The model then uses this information to calculate the overall cost (Equation \ref{eq:cost-formula}) and uses that cost to choose a \pstate{} such that the cost would be minimum. Figure \ref{fig:model-high-level} depicts all the steps performed by \sysname{} at a high level. \sysname{} takes a binary (target application) as input and generates memory traces by profiling the application using MemGaze. \sysname{} then uses these traces (along with the original binary) to perform footprint growth analysis (\S\ref{sec:work-char}) and candidate selection (\S\ref{sec:selecting-code-regions}) to generate code regions along with target \pstates{}. Finally, \sysname{} uses this information to inject target \pstate{} calls for each specified code region in the original binary.

The model is broken into two pieces: (1) selecting the memory intensive code regions of the application, and (2) trade-off decision, i.e., placing the \pstate{} transition calls in appropriate locations in the program such that the transition overhead is amortized over time.

Available parameters to the model are described in Table~\ref{tab:terminology}.

\begin{table*}[t]
\centering
\caption{Terminology and Description}
\label{tab:terminology}
\setlength{\tabcolsep}{4pt}
\begin{tabular}{>{\raggedright\arraybackslash}p{0.15\linewidth} >{\raggedright\arraybackslash}p{0.06\linewidth} >{\raggedright\arraybackslash}p{0.73\linewidth}}
\hline
\textbf{Term} & \textbf{Symbol} & \textbf{Description} \\
\hline
Footprint Growth Rate & $F$ & Average footprint (unique memory accesses) per memory access of a program, computed per code region. Classified into three subcategories: \\
\quad Irregular       & $F_{\text{irr}}$ & Rate of accesses made in an irregular pattern, e.g., pointer chasing. \\
\quad Strided         & $F_{\text{str}}$ & Rate of accesses separated by a distance $k$ (where $k \geq 1$). \\
\quad Constant        & $F_{\text{con}}$ & Rate of accesses to scalar data within a stack frame or to global data. \\
Code Region           & CR        & A subsection of the program represented as a set of basic blocks, annotated with footprint growth rate. \\
Calling Context Tree  & CCT       & Constructed from \textit{last branch record} (LBR) events captured during memory trace collection by MemGaze. It is incomplete as it lacks invocation counts, which cannot be inferred from sampled LBR data alone. \\
Transition Cost       & $P$       & Time required to change from the current to the desired P-state. \\
Level                 & $L$       & Call depth of the target function (or code region) in a CCT. \\
Cycles                & $T$       & Total time (in cycles) spent executing a function, excluding time in child functions (exclusive metric). \\
\hline
\end{tabular}
\end{table*}

Given these inputs we describe our \pstate{} instrumentation cost metric (C):
\begin{equation}
\label{eq:cost-formula}
C = \frac{P \times L}{T \times F_{\text{irr}}}
\end{equation}

\subsection{Selecting Memory Intensive Sequences}
\label{sec:selecting-code-regions}

To identify memory intensive regions, we parse the application's memory trace obtained from MemGaze, which gives us fine-grained access information. A memory trace from MemGaze is organized as time-series data where each trace element has an \textit{instruction pointer} of the LOAD instruction, the \textit{basic block} that contains the IP, the \textit{memory address} that was accessed, and \textit{time}. We use this trace to construct a weighted, directed graph that compresses the trace (which is total-order time series data) into a representation of the most frequent dynamic basic-block transitions. In the graph, each node represents a basic block, edges indicate a ``followed by'' relationship (i.e., access from source block occurred before access from destination block in the trace) and edge weight represents access intensity which is a count of all such occurrences in the trace. Constructing the graph at basic block granularity captures sequencing while avoiding unnecessarily large graphs. 

\begin{figure}
    \centering
    \includegraphics[width=\linewidth]{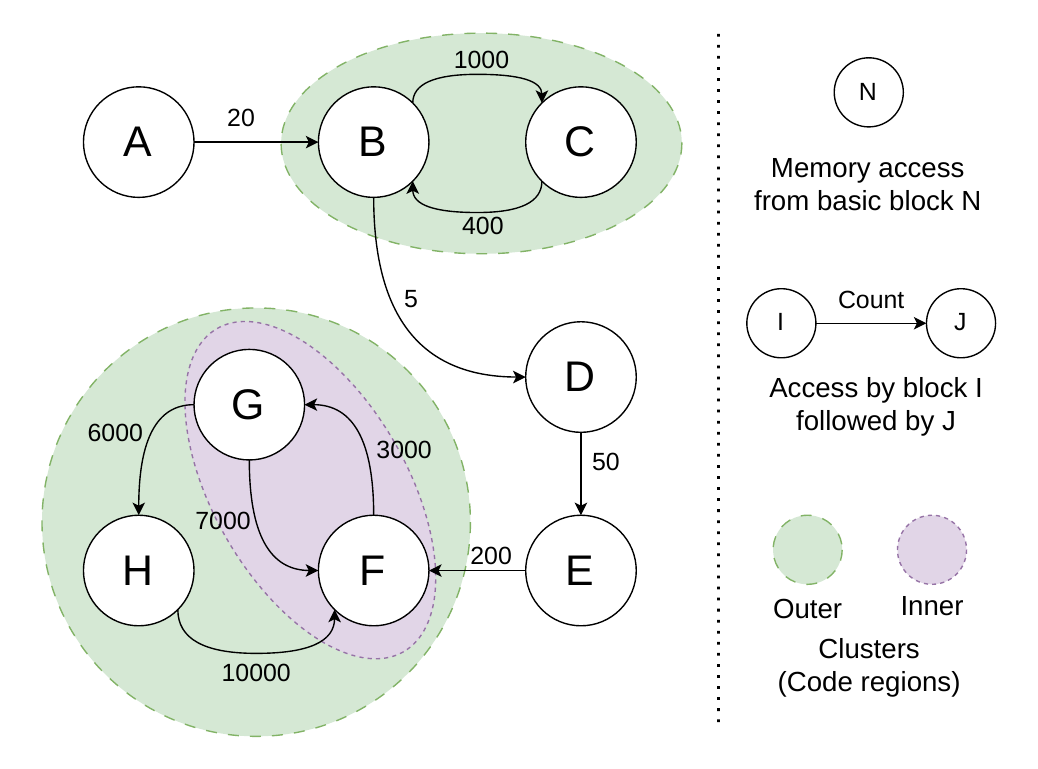}
    \caption{Clustering basic blocks to form code regions based on the memory access trace.}
    \label{fig:clustering}
\end{figure}

We cluster the graph (using the Louvain
Method~\cite{Ghosh:2018:PDPS:LouvainMethod}) to identify all sub-graphs that
are tightly connected to each other, resulting in clusters of basic-block
sequences we call \textit{code regions}. Each code region is attributed with
its \textit{footprint growth rate} along with its \textit{access type} (\textbf{Goal
G\ref{goal:irreg}}). To explain this visually, Figure~\ref{fig:clustering}
depicts a graph where edges from one node (basic block) to another represent
access intensity and clustered sub-graphs represent code regions. Moreover,
since the Louvain method produces clusters with proper hierarchical relationships,
we can identify inner vs. outer sequences, which lets us select code-region
boundaries that amortize \pstate{} transition cost.

Given these clusters (code regions), we need a way to filter out all those code regions that are not beneficial to execute at a lower \pstate{}, i.e., code regions with relatively low irregular memory access footprint. Code regions with regular access patterns (constant or strided) have better locality than irregular ones and are therefore poor candidates for execution at a lower \pstate{}.. To do this we construct a new metric, \textit{Relative Footprint Growth} in Equation \ref{eq:relative_F_irr}, that can be used to identify the code regions with low irregular footprint growth rate and filter them out of the candidate code regions for further instrumentation.
 
The relative footprint growth metric for the $F_{\text{irr}}$ can be calculated as follows:
\begin{equation}
    \text{Relative } F_{\text{irr}} = \frac{F_{\text{irr}}}{F_{\text{irr}} + F_{\text{str}} + F_{\text{con}}}
    \label{eq:relative_F_irr}
\end{equation}

\subsection{Cost-benefit Analysis}
\label{sec:cost-model}
As part of the model we need to compute the cost of choosing an effective \pstate{} for a code region. Since every node in the $CCT$ is attributed with a code region, we compute this cost for each node using Equation~\ref{eq:cost-formula}. To map a cost to an effective \pstate{}, we define cost thresholds such that a \pstate{} is chosen if the computed cost falls within a particular threshold range for a node in $CCT$. We construct these
threshold ranges empirically based on the benchmarks we evaluated. For example, if the computed cost is within a range (900 to 1000), then we choose \pstate{} 30. If the cost is higher than all available cost thresholds at a given node, we skip that node for instrumentation, and move to its parent. This is repeated until we reach the root node. We essentially end up with code regions with corresponding efficient \pstate{} transition calls
(\textbf{Goal G\ref{goal:balance}}).

\section{Implementation}
\label{sec:implementation}

In this section, we describe the implementation details of \sysname{} and its subcomponents: the cost model implementation (\S\ref{sec:model-impl}), a userspace API along with a kernel driver to enable manual \pstate{} transition (\S\ref{sec:pstate-impl}), and a binary instrumentation tool to inject \pstate{} transition calls into the target binary (\S\ref{sec:inject-pstate}).

\begin{figure}
    \centering
    \includegraphics[width=0.9\columnwidth]{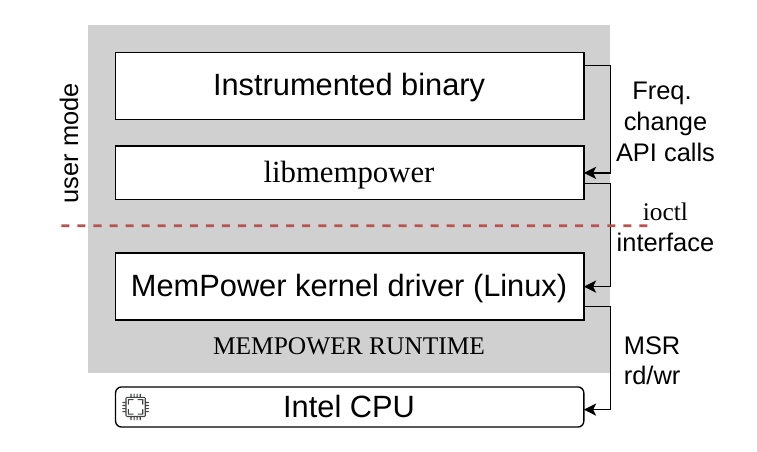}
    \caption{\sysname{} Architecture.}
    \label{fig:implementation}
\end{figure}

\subsection{Model Implementation}
\label{sec:model-impl}
We implement \sysname{}'s model (\S\ref{sec:model}) as a Python framework that takes memory traces and LBR data (which are generated by MemGaze) as input and outputs the code regions with corresponding \pstate{} transitions. 

Initially, \sysname{} invokes a graph clustering tool (implemented in C) to analyze the input memory trace and generate candidate code regions as described in \S\ref{sec:selecting-code-regions}. Then, a Python script takes the LBR data and the generated code regions as input to construct a CCT where each node represents a candidate code region. Finally, \sysname{}'s cost benefit model (\S\ref{sec:cost-model}), implemented in Python, iterates over this CCT to select efficient \pstates{} for each code region. Thus, the final output of \sysname{}'s model is a set of code regions (i.e., start and end instruction pointers) along with the predicted \pstate{} value for each region. 

\subsection{Userspace Library and Kernel Driver}
\label{sec:pstate-impl}

\begin{table}
\centering
\caption{\sysname{} userspace API.}
\label{tab:userspace-api}
 \resizebox{\columnwidth}{!}{
\begin{tabular}{@{}ll@{}}
\toprule
\textbf{API call}    & \textbf{Description}                \\
\midrule
\texttt{mempow\_init()}       & Initialize \sysname{}    \\
\texttt{mp\_set\_pstate\_min()}   & Transition to the lowest \pstate{}    \\
\texttt{mp\_set\_pstate\_max()}   & Transition to the highest \pstate{}   \\
\texttt{mp\_set\_pstate\_eff(int p)}   & Transition to the given \pstate{} \\
\texttt{mempow\_deinit()}       & Tear down \sysname{} \\
\bottomrule
\end{tabular}
}

\end{table}

We implement a light-weight userspace library, specified in Table~\ref{tab:userspace-api}, that interfaces with our custom Linux driver  to perform on-demand \pstate{} transition calls while running a program.

The \verb|mempow_init()| and \verb|mempow_deinit()| functions perform setup and cleanup, respectively, coordinating with the custom kernel driver to perform required \pstate{} transitions. The \verb|mp_set_pstate_min()|, \verb|mp_set_pstate_max()| and \verb|mp_set_pstate_eff()| functions implement the \textit{minimum}, \textit{maximum}, and \textit{effective} \pstate{} transition functionality respectively. Here, \textit{minimum} means the lowest available \pstate{}, \textit{maximum} means the highest available \pstate{}, and \textit{effective} means the predicted \pstate{} (output of the model described in \S\ref{sec:model}).

We take inspiration from TURBO Diaries~\cite{Wamhoff:2014:ATC:TURBO} work to implement a light-weight Linux kernel module to perform \pstate{} transitions on demand. A \pstate{} transition can be triggered by writing the desired \pstate{} value to \verb|IA32_PERF_CTL| MSR (Model Specific Register) on the desired CPU core. We implement this in a kernel driver that exposes an \verb|ioctl()| interface to userspace; the driver takes the desired \pstate{} as input and performs the transition by writing that value to the executing core's \verb|IA32_PERF_CTL| MSR.

\subsection{Binary Instrumentation}
\label{sec:inject-pstate}

We instrument appropriate \pstate{} transition calls defined in \S\ref{sec:pstate-impl}. This can be done in two ways: at the source level (if source is available) or at the binary level. We use static binary instrumentation to evaluate all the benchmarks for our work to demonstrate that using \sysname{} does not require programmers to recompile their code (\textbf{Goal G\ref{goal:low-burden}}). However, if a developer for any reason prefers to insert P-state transition calls directly into the code, they can do so using our userspace library, which is also documented in Table \ref{tab:userspace-api}.

For both cases, the injection process is as follows:
\begin{enumerate}
 \item Call the \verb|mempow_init()| function before (or at the very beginning of) the \verb|main()| function to initialize the library.
    \item For each candidate code region:
    \begin{enumerate}
        \item Call the \verb|mp_set_pstate_eff()| function before (or at the very beginning of) the candidate region to perform \textit{effective} \pstate{} transition.
        \item Call the \verb|mp_set_pstate_max()| function after (or at the very end of) the candidate region to perform \textit{maximum} \pstate{} transition.
    \end{enumerate}
    \item Call the \verb|mempow_deinit()| function after (or at the very end of) the \verb|main()| function to perform library cleanup.
\end{enumerate}

In Step 2.(b), the \pstate{} is reset from \textit{effective} frequency to \textit{maximum} frequency such that the portions of the program that are not part of candidate functions can be executed at \textit{maximum} frequency as they are filtered out by the selection process in \S\ref{sec:selecting-code-regions}.

\begin{figure}
\centering
\begin{lstlisting}[style=ics-mono, escapechar=!]
// in main
mempow_init(); 
mp_set_pstate_eff(p); // manual P-State ctrl
// ...
mp_set_pstate_max(); // return to max freq
mempow_deinit();
\end{lstlisting}
\caption{Code snippet demonstrating our C userspace API.}
\label{code:ex-inject}
\end{figure}

Figure~\ref{code:ex-inject} illustrates how application code is modified after injecting userspace library function calls. Lines 2, 3, 5, and 6 represent the appropriate userspace library calls that are injected into a target application after performing the aforementioned injection process. Line 3 is where the effective \pstate{} transition is performed.

In the case of \sysname{}, we implement a DynInst based binary instrumentation tool (written in C++) to inject appropriate \pstate{} transition calls directly into the target binary based on the output of \sysname{}'s model component (\S\ref{sec:model-impl}).

\boldpara{Availability} Source code for \sysname{} is available at \url{https://github.com/nandavelugoti/MEMPOWER}.

%============================================================================
%============================================================================

\section{Evaluation}
\label{sec:evaluation}

In evaluating \sysname{}, we seek to answer the following questions:

\begin{itemize}

\item How effective is attributing footprint metrics to a code region to make better choices for specific \pstates{} relative to the HW/firmware? (\S\ref{sec:miniVite})
% A. Show the co-relation study done for \pstate{} and FP growth rate along with the frequency distributions

\item To what degree does \sysname{} capitalize on memory intensive workloads, and how does it handle compute intensive workloads? (\S\ref{sec:NAS-HPCG})

\item How does \sysname{} compare against the default mechanisms across all workloads? (\S\ref{sec:mempwr-eval})

\item How does each component of \sysname{} contribute to overall application performance? (\S\ref{sec:ablation})

\end{itemize}

\begin{table}[]
\centering
\caption{Experiment setup.}
\label{table:exp-setup}
\begin{tabular}{@{}ll@{}}
\toprule
    \textbf{Component}    & \textbf{Description}                \\
\midrule
    CPU & 12th Gen Intel\textregistered{} Core\texttrademark{} Processors\\
    Number of Cores&16 (8 P-cores and 8 E-cores)\\ 
    DRAM & 128 GB \\ 
    OS & Linux 6.0 \\
    Frequency driver & \texttt{acpi-cpufreq}  \\ 
    Default policy & \texttt{schedutil} \\
    \sysname{} policy & \texttt{userspace} + \sysname{} driver \\
\bottomrule
\end{tabular}
\end{table}

\boldpara{Experimental setup} 
We evaluate our experiments on the system specified in Table \ref{table:exp-setup}. We compile our OpenMP-based benchmarks with GCC (v11.5). We also use MemGaze (v1.0) for profiling memory behavior of the benchmarks. We run the benchmarks with 8 threads and pin the OpenMP threads to performance cores using the \verb|OMP_NUM_THREADS| and \verb|OMP_PROC_BIND| environment variables. Finally, we run each benchmark 10 times and use the arithmetic mean of our measurements to present data.

\begin{table}
\centering
\caption{Breakdown of miniVite's code regions (attributed to functions) along with the constant, strided and irregular memory accesses.}
\label{tab:minivite-breakdown}
\resizebox{\columnwidth}{!}{
\begin{tabular}{lrrrr}
\toprule
\textbf{Code Region}
& $F_{\text{con}}$
& \textbf{$F_{\text{str}}$}
& \textbf{$F_{\text{irr}}$}
& \textbf{P State} \\
\midrule
\texttt{distUpdateLocalCinfo}     & 0.00 & 0.60 & 0.60 & 30 \\
\texttt{fillRemoteCommunities}    & 0.00 & 0.40 & 0.05 & 32 \\
\texttt{distSumVertexDegree}      & 0.00 & 0.41 & 0.62 & 30 \\
\texttt{distBuildLocalMapCounter} & 0.90 & 0.39 & 0.78 & 29 \\
\bottomrule
\end{tabular}
}

\end{table}

\subsection{Memory Behavior vs. \pstate{}}
\label{sec:miniVite}
In order to understand the effectiveness of using memory metrics (footprint growth and access class) to pick an efficient \pstate{} for the memory intensive code region of a workload, Table \ref{tab:minivite-breakdown} shows the \pstate{} picked by \sysname{} (process detailed in \S\ref{sec:cost-model}) for each memory intensive code region (\texttt{distUpdateLocalCinfo}, \texttt{fillRemoteCommunities}, \texttt{distSumVertexDegree}, and \texttt{distBuildLocalMapCounter}) of \linebreak \textit{miniVite} along with the corresponding memory metrics ($F_{\text{con}}$, $F_{\text{irr}}$, and $F_{\text{str}}$). For the region with highest $F_{\text{irr}}$, \verb|distBuildLocalMapCounter|, \sysname{} chose \pstate{} 29 (third highest \pstate{}), and for the region with lowest $F_{irr}$, \verb|fillRemoteCommunities|, \pstate{} 32 (maximum) is chosen. Figure \ref{fig:motivation} shows the effectiveness of this policy by comparing \sysname{}'s relative EDP with the baseline (\verb|schedutil|) setting and the manual selection of all possible \pstates{}.

\subsection{Effectiveness of \sysname{}}
\label{sec:NAS-HPCG}

\begin{figure*}
    \centering
    \includegraphics[width=\textwidth]{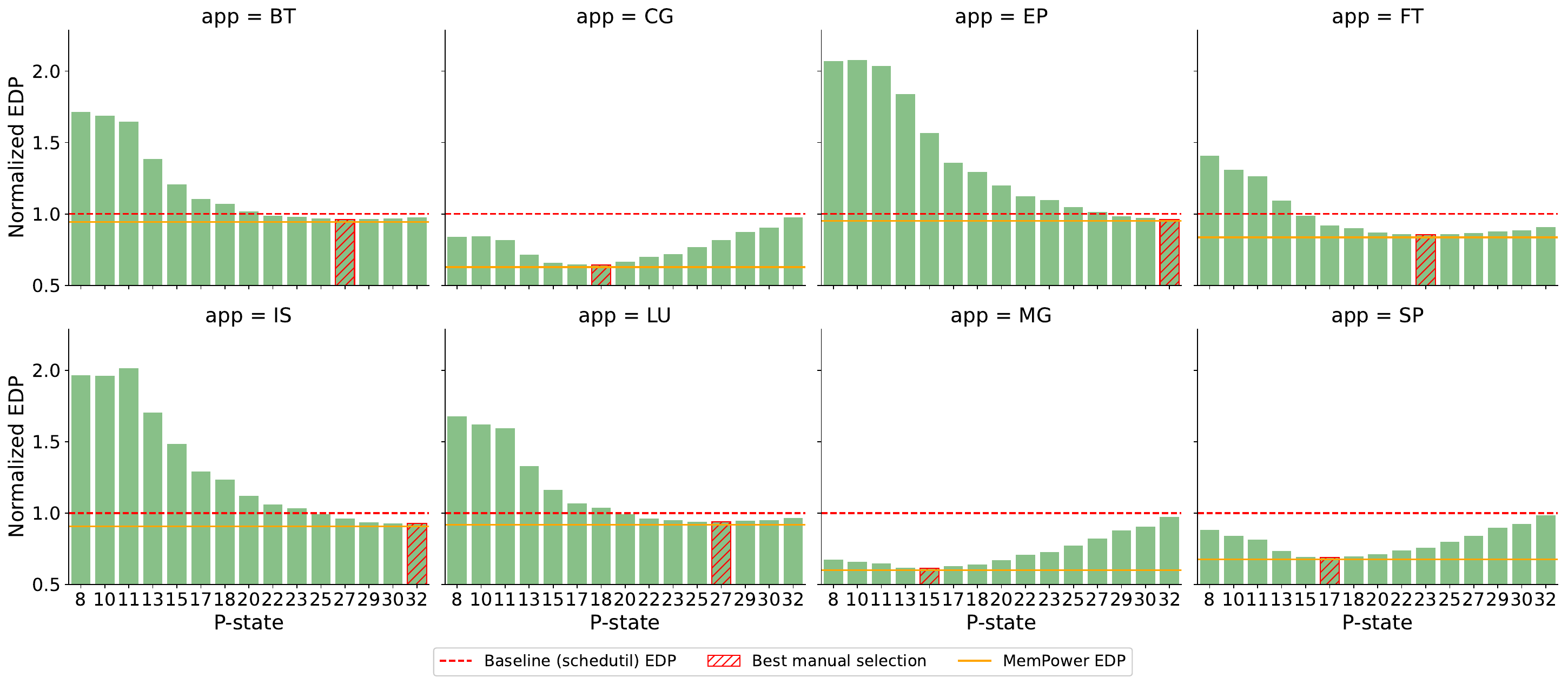}
    \caption{Relative EDP of NAS Parallel Benchmarks when executed at all \pstate{} configurations. Highlighted bars show where manually picked frequencies do better than the baseline. Lower is better.}
    \label{fig:nbp-edp}
\end{figure*}

To understand the degree to which \sysname{} capitalizes on the opportunity for
picking an efficient \pstate{} based on the memory characteristics of
a given workload, we evaluate NAS Parallel Benchmark~\cite{BAILEY:1991:NAS-PB}
and HPCG~\cite{DONGARRA:HPCG:2013} which contain workloads with varying degrees
of memory/compute intensity and access patterns, and we discuss the impact on
overall EDP for each workload over baseline and \sysname{}.

Figure \ref{fig:nbp-edp} shows the normalized EDP of the NAS parallel benchmark suite scaling across all available \pstates{} when compared to the baseline (\verb|schedutil|). The workload size is constant across all tests, which is Class B (average execution time is around 60 to 70 seconds).

% best
CG, MG, and SP show the best results with 40\% to 42\% reduction of overall EDP when compared to the baseline. This is because all the applications have large memory intensive sections: \verb|conj_grad| in CG; \verb|norm2u3| and \verb|mg3p| in MG; and \verb|x_solve|, \verb|y_solve| and \verb|z_solve| in SP. Thus, when the \pstates{} are lowered for these regions, the cost of transition overhead amortizes over time and \sysname{} capitalized on this opportunity to reduce the EDP.

% moderate
In the case of FT, even though it has a significant number of irregular memory intensive sections, it also has regions with  intense floating point operations which can be run at higher \pstates{} to reduce the execution time. Even so, \sysname{} was able to reduce relative EDP by 20\%. This is because for code regions such as these, where operational intensity is high, \sysname{}'s candidate selection process will not tag these regions as candidates for effective \pstate{} instrumentation. Therefore, \sysname{} sets the \pstate{} to the maximum value for these code regions, thus reducing the overall execution time (i.e., reduction in EDP). Similarly, IS also has irregular memory access regions resulting in a 10\% reduction in EDP. Both BT and LU are memory intensive with strided memory access patterns resulting in better locality;  using \sysname{} reduced the relative EDP by 10-15\% compared to the baseline.

% poor
EP has only 6\% reduction in EDP compared to the baseline. This is because EP is a highly parallel and compute intensive workload. Since \sysname{} recognizes that EP does not have memory intensive code regions with irregular accesses (\S\ref{sec:selecting-code-regions}), it sets the \pstate{} to the maximum available frequency by default. This means running the workload at the max available \pstate{} is the best option.

\begin{figure}
    \centering
    \includegraphics[width=0.8\columnwidth]{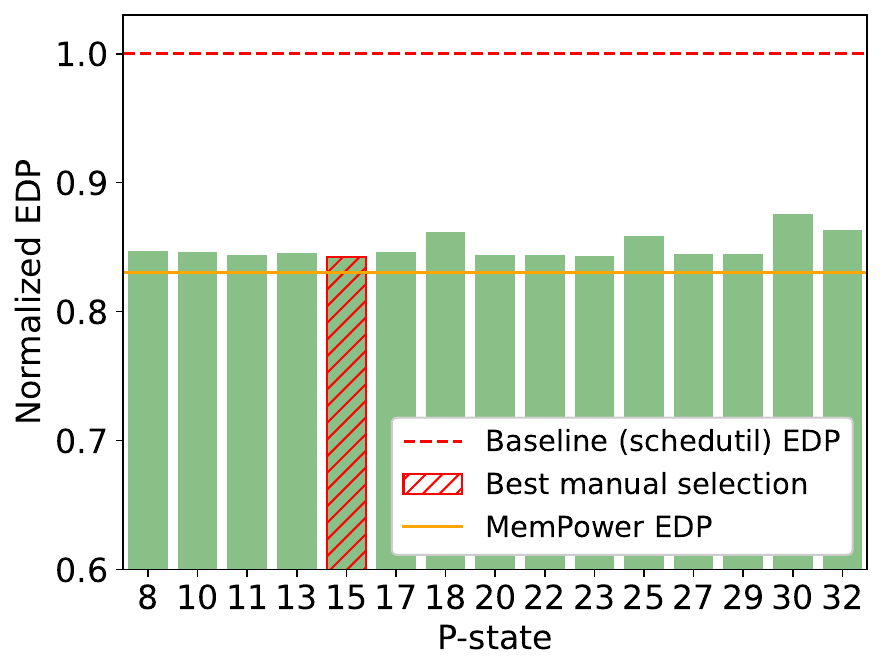}
    \caption{Normalized EDP of HPCG when executed at all \pstate{} configurations. Lower is better.}
    \label{fig:hpcg}
\end{figure}

In addition to NAS Parallel benchmarks, we also evaluate HPCG, in Figure \ref{fig:hpcg}, to showcase
\sysname{}'s ability to capitalize on sparse and irregular memory behavior to
pick an efficient \pstate{} resulting in an 18\% reduction in EDP relative to
the baseline. This is because HPCG's workload consists of four
major subroutines: sparse matrix-vector multiplication (SpMV), symmetric
Gauss-Seidel (SymGS), vector updates (WAXPBY) and dot product (DDOT), where
SpMV has irregular memory access pattern, SymGS has a recursive access pattern
(i.e., current iteration compute depends on data from the previous
iteration) and the rest have streaming/strided access patterns. Thus \sysname{}
chose lower \pstate{} (29) for SpMV, SymGS and maximum value (32) for WAXPBY
and DDOT. The inputs to HPCG are: $nx=104, ny=104, nz=104$, and $rt=60$ (seconds).

In summary, \sysname{} was able to capture the memory characteristics for a given workload and choose an efficient \pstate{} scheme, resulting in
reduction in overall application EDP, of up to 42\%, compared to the baseline.
Furthermore, in all the NAS workloads, \sysname{} performed as well as or
better than the best manual selection.

\subsection{\sysname{} vs. Other Policies}
\label{sec:mempwr-eval}

\begin{figure}
\centering
\includegraphics[width=1.0\columnwidth]{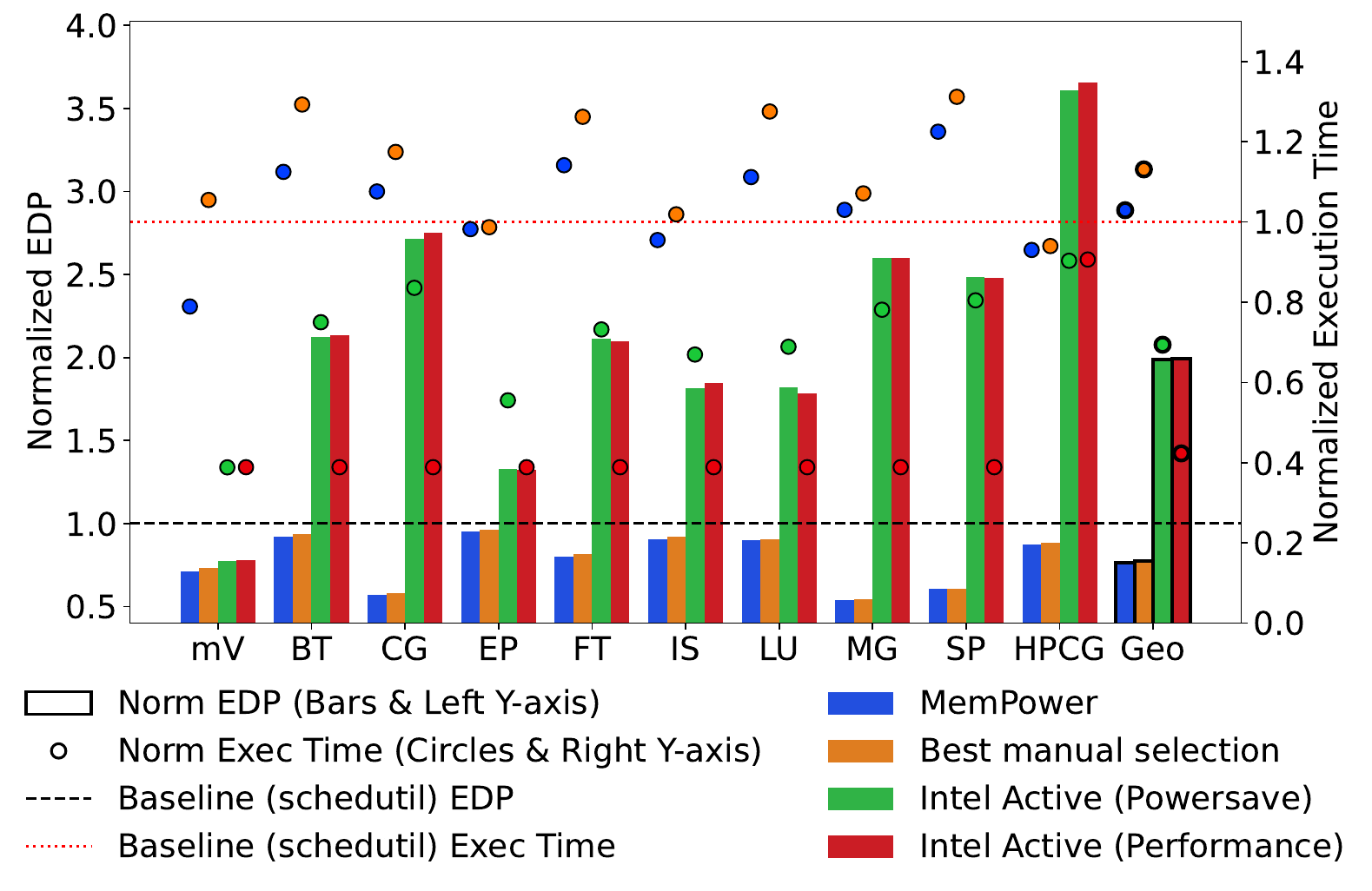}
\caption{Comparing normalized EDP and execution times of \sysname{} vs. Intel active mode vs. schedutil vs. best manual \pstate{} selection policies across all benchmarks. The last set of bars and circles are the geometric means (EDP and execution time) of all benchmarks. Lower is better.}
\label{fig:all-bench}
\end{figure}

We evaluated \sysname{} on the following benchmarks: NAS Parallel Benchmarks, HPCG (\S\ref{sec:NAS-HPCG}), and miniVite (\S\ref{sec:miniVite}). These workloads contain different data structures: graphs in \textit{miniVite}, arrays in NAS parallel benchmark's \textit{IS}, and matrices in rest of the workloads. Furthermore, these workloads also contain different memory access patterns, i.e., \textit{irregular} accesses in \textit{miniVite}, \textit{IS}, and \textit{CG}; \textit{strided} accesses in \textit{FT}, \textit{LU}, \textit{MG}, and \textit{SP}; \textit{constant} accesses in \textit{EP}. Figure~\ref{fig:all-bench} summarizes the best cases of energy utilization and time costs resulting from our model when compared to the OS (\verb|schedutil|) baseline, with 6\% to 42\% reduction in overall EDP (bars with left Y-axis). The geometric mean of all the benchmarks (last set of bars in Figure~\ref{fig:all-bench}) is a 20\% reduction in EDP. We observe that in all cases, \sysname{} meets or exceeds the best manual selection (\textbf{Goal G\ref{goal:edp}}).

In addition to \verb|schedutil| and manual \pstate{} selection policies, we also compare \sysname{} with 
Intel active mode policy which, unlike \verb|schedutil|, has the capability to use boost/turbo frequency ranges (3.3GHz to 6GHz) but lacks the capability of choosing a static \pstate{} (one can only set upper and lower bounds). Intel active mode has two governors: \textit{performance} and \textit{powersave}. In both cases, \sysname{} outperforms them by $2\times$ in terms of EDP, because active mode's use of turbo frequencies increases overall energy consumption significantly even though it lowers execution time relative to the other policies.

Since EDP is a combined metric (product of energy and time), it can mask the performance degradation of an application while selecting a lower \pstate{}. To address this concern, we have included the normalized execution times (circles with right Y-axis) for all policies across all benchmarks. We can see that the performance degradation for \sysname{} policy across all benchmarks is less than 3\% (geometric mean).

\subsection{\sysname{} Component Analysis}
\label{sec:ablation}
\begin{figure}
\centering
\includegraphics[width=1.0\columnwidth]{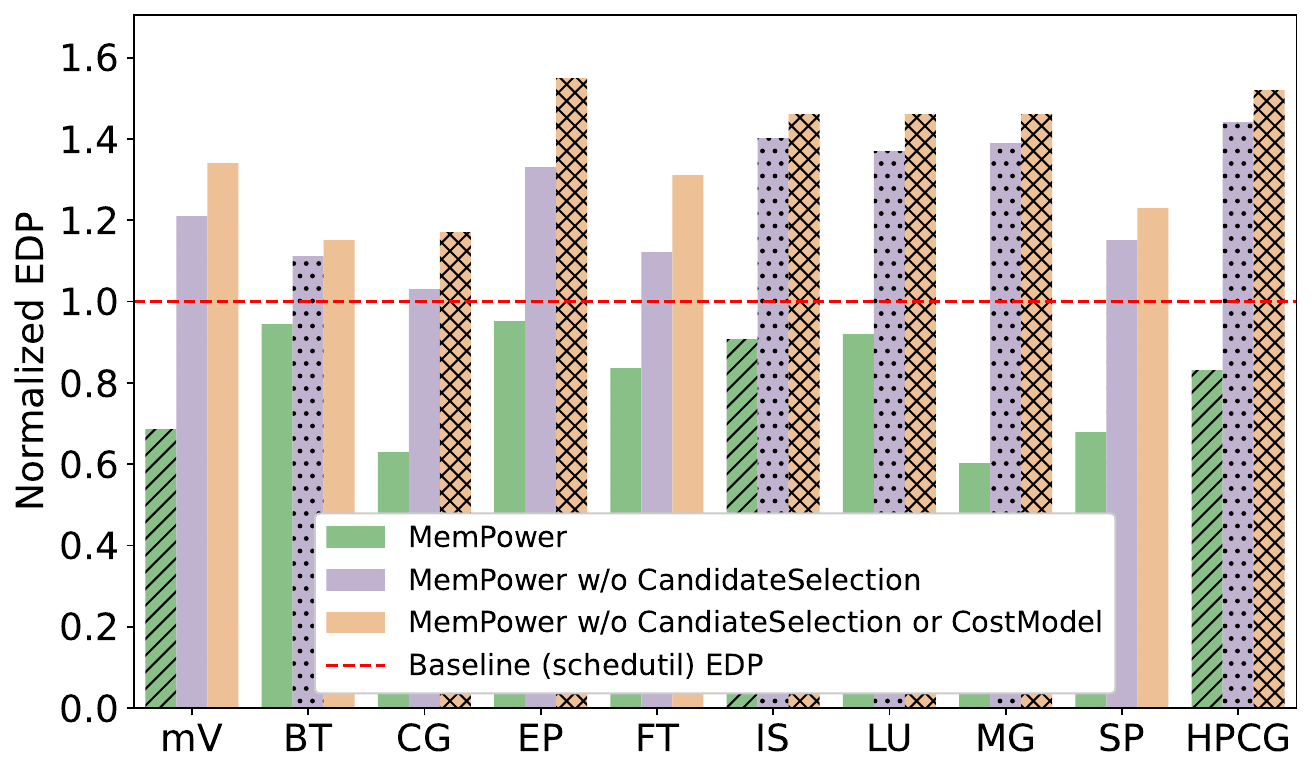}
\caption{Evaluating the EDP of \sysname{} along with its components enabled/disabled with respect to the baseline. Lower is better.}
\label{fig:ablation}
%\vspace{-0.2in} %% JBMF: Delete if more space can be found
\end{figure}

To explore the effectiveness of \sysname{}'s components, we evaluate \sysname{} on all benchmarks with various configurations: 1) \sysname{} without the candidate selection; 2) \sysname{} without the candidate selection and the cost model; and 3) \sysname{} without any change. Figure \ref{fig:ablation} shows the normalized EDP of all these configurations relative to the baseline, and we can observe that turning any of the \sysname{}'s subcomponents off would result in poor performance compared to \sysname{} as a whole. For configuration 1, the overall EDP is worse than \sysname{}'s because the code regions that are supposed to be filtered out by the candidate selection process instead qualify for instrumentation, increasing the overall overhead due to \pstate{} transitions. Similarly, for configuration 2, where both candidate selection and the cost model are turned off (i.e., \pstates{} are chosen manually), the EDP is much worse than that of \sysname{} because of the increased \pstate{} transition overhead along with inefficient placement of those transition calls. One other possible configuration, \sysname{} with candidate selection but without the cost model, is an invalid configuration, because the cost model is what assigns a \pstate{} to each selected region.

\section{Discussion}
\label{sec:discussion}

\boldpara{Key takeaways} 1) hardware is not aware of high-level memory characteristics, like footprint growth, which opens up opportunities for efficient power management at the software level and 2) amortizing the cost of a \pstate{} transition is key in order to achieve efficient EDP.

\boldpara{Improving \sysname{}}
Intel's hardware analyzes the instruction mix of a running program to choose a P-state. However, based on the observations from \sysname{}, we speculate that incorporating auxiliary information such as cache misses collected at runtime (using the existing performance counters such as \linebreak \verb|LONGEST_LAT_CACHE.MISS| and \verb|LLC_MISSES|) in addition to instruction mix would significantly improve hardware's ability to choose an efficient P-state as it would allow the hardware to recognize latency-bound memory behavior. Furthermore, since \sysname{} is a framework as well as a tool, it can be extended to incorporate other memory metrics that can capture spatial locality, memory contention and any other memory related metric to achieve desired performance-energy trade-off. Another aspect of improving \sysname{} is to enhance the cost model itself by either incorporating other memory metrics to the cost formula (Equation~\ref{eq:cost-formula}) or replacing it entirely. In addition, a programmer can configure a custom cost threshold policy to the cost model depending on their needs.

\boldpara{Developer impact} Even though \sysname{} does not require code
recompilation and is application agnostic, it requires a programmer to profile
and instrument the target application binary. For the benchmarks used in this
paper, \sysname{} typically takes three to five minutes of developer time for
each individual application. Once done, future runs do not require any
instrumentation for the same input. Furthermore, the system administrator needs
to install the \sysname{}'s kernel driver. Making \sysname{} completely
transparent to the user and integrating it at the system level (OS or hardware)
would encourage wide-spread adoption. 

\boldpara{Limitations} A key limitation of \sysname{} is that it 
currently works only with specific hardware, namely Intel x86 CPUs with \verb|PT_WRITE| support, as it is built on top of MemGaze.
In terms of other architectures like ARM, MemGaze can be
modified to support MMIO-based STM (System Trace Macrocell) or the ITM
(Instrumentation Trace Macrocell) interface to collect the memory traces.

\boldpara{OpenMP workloads} All the evaluations
(\S\ref{sec:evaluation}) of \sysname{} are done on OpenMP enabled HPC workloads
as we are trying to optimize EDP at the node level. Though out of scope for our paper, it would be interesting to study
\sysname{} on workloads at scale using MPI and we consider this to be future work.

\section{Related work}
\label{sec:related-work}

At the hardware level, DVFS~\cite{WEISER:1994:DVFS} enables dynamic scaling of voltage
and frequency in CPU cores. Kumar et al.~\cite{KUMAR:2003:MICRO:hetero-multicore-power}
introduced single-ISA heterogeneous multicore architectures---the basis for today's
P-core/E-core designs---and later showed that performance can be improved~\cite{KUMAR:2004:ISCA:hetero-multicore-workload}. 
At the OS level, there are 
workload scheduling policies such as scheduling on ``wimpy'' cores~\cite{HRUBY:2013:ATC:slow-is-fast}, 
PIE~\cite{VAN:2012:ISCA:PIE}, POW~\cite{ELLSWORTH:2015:POW}, and PShifter~\cite{GHOLKAR:2018:PSHIFTER}. At the compiler level, many works have proposed algorithms~\cite{HSU:2003:PLDI:compiler-algorithm}, and models/frameworks such as 
DAE~\cite{KOUKOS:2013:DAE-DVFS}, PULSE~\cite{TANG:2025:ASPLOS:PULSE} that focused on reducing application power utilization. At the library/runtime level, runtime systems such as PART~\cite{HSU:2005:POWER-RUNTIME}, Adagio~\cite{ROUNTREE:2007:MPI-ENERGY, ROUNTREE:2009:ADAGIO}, COUNTDOWN~\cite{Cesarini:2018:ANDRE:COUNTDOWN}, UPSCavenger~\cite{GHOLKAR:2019:UNCORE}, and online predictors~\cite{CURTIS-MAURY:2006:POWPERF} were proposed to balance performance and energy usage.

At the application level, similar to \sysname{}, Magklis et
al.~\cite{MAGLIKIS:2003:PROFILE-DVFS} proposed a profile-driven approach where
an application is profiled to identify code regions for which
efficient power transitions are applied. However, there are a few key differences when
compared to \sysname{}: 1) their approach profiles an application over multiple
``training'' runs whereas \sysname{} requires the application to be run once to
obtain the memory trace; 2) their approach characterizes code regions as L+F
(Loops and Functions) which are coarse grained whereas \sysname{} code regions
can be more fine-grained, i.e., code blocks with arbitrary start and end IP addresses; and 3) their
approach is evaluated on a simulator whereas \sysname{} is evaluated on real
hardware. Similarly, PowerDial~\cite{HOFFMANN:2011:DYN-KNOBS} augments a target 
application and manages power using a heartbeat based power control system.
Another work used customized JIT runtime to improve energy efficiency of an 
application~\cite{WU:2005:JIT-DVFS}. Unlike \sysname{}, these works 
require an online component (apart from an instrumented binary) to manage power.

Apart from the works mentioned above, there are others that have used memory
characterization but not in the context of power/energy optimization. For
example, using miss ratio curves (MRCs) to characterize storage
workloads~\cite{WIRES:2014:OSDI:storage-counter-stacks}, custom page
replacement policies for memory intensive
applications~\cite{WU:2024:IISWC:PAGE-REPLACE}, using memory footprint to
improve LLM fine-tuning~\cite{WANG:2025:ATC:JENGA, HUANG:2025:ATC:mTuner}, and
characterizing memory behavior of Google's TCMalloc memory allocator at
warehouse scale~\cite{ZHOU:2014:ASPLOS:CHARACTERIZE-TCMalloc}. In contrast to
all the studies referenced in this section, \sysname{} is a novel,
profile-driven, model-based power management framework that improves
applications' performance per Watt.

\section{Conclusion}
\label{sec:conclusion}

We presented \sysname{}, a novel, memory-centric modeling framework that
performs profile-based analysis to identify and actuate effective DVFS
transitions for HPC workloads. We demonstrated that \sysname{} can save up to
42\% in total EDP compared to the default OS/hardware coordinated mechanism.
Our approach imposes minimal burden on users, and in
principle can be applied to a broader set of workloads. 
In future work, we plan to extend our model further by incorporating other
memory metrics such as reuse distance or MPKI (misses per kilo-instruction) and
further investigate the EDP trade-off in MPI enable applications. We also envision the development of more
sophisticated governors for the kernel. We plan to extend our memory
characterization model to improve aspects such as deciding when to offload
memory intensive tasks, e.g., in architectures that employ near-data
processing. 

\bibliographystyle{IEEEtran}
\bibliography{refs,kyle}
\end{document}